\documentclass[12pt]{iopart}

\begin{document}

\title{Treatment of like-particle pairing with quartets}

\author{M. Sambataro}
\address{Istituto Nazionale di Fisica Nucleare - Sezione di Catania\\
Via S. Sofia 64, I-95123 Catania, Italy}
\ead{michelangelo.sambataro@ct.infn.it}

\author{N. Sandulescu}
\address{National Institute of Physics and Nuclear Engineering,\\
P.O. Box MG-6, 76900 Bucharest-Magurele, Romania}
\ead{sandulescu@theory.nipne.ro}

\begin{abstract}
The ground state correlations induced by a general pairing Hamiltonian in a finite system of like fermions are described in terms of four-body correlated structures (quartets). These are
real superpositions of products of two pairs of particles in time-reversed states. 
Quartets are determined variationally through an iterative sequence of  diagonalizations of the Hamiltonian in restricted model spaces and are, in principle, all distinct from one another.
The ground state is represented as a product of quartets to which, depending on the number of particles (supposed to be even, in any case), an extra collective pair is added. The extra pair is also determined variationally. 
In case of pairing in a spherically symmetric mean field, both the quartets and the extra pair (if any) are characterized by a total angular momentum $J=0$. Realistic applications of the quartet formalism are carried out for the Sn isotopes with the valence neutrons in the 50-82 neutron shell. Exact ground state correlation energies, occupation numbers and pair transfer matrix elements are reproduced to a very high degree of precision. The formalism also lends itself to a straightforward and accurate description of the lowest seniority 0 and 2 excited states of the pairing Hamiltonian. A simplified representation of the ground state as a product of identical quartets is eventually discussed and found to improve considerably upon the more traditional particle-number projected-BCS approach.
\end{abstract}

\pacs{21.60.-n,74.20.-z,74.20.Fg}

\maketitle

\section{Introduction}
Pairing strongly affects the behavior of a large variety of quantum many-body systems ranging from very small, like nuclei, to very large, like stars \cite{dean}. The concept of pairing made its first appearance in the condensed matter context in an attempt to explain the phenomenon of electron superconductivity in metals \cite{cooper,bcs}. Soon after, this concept was imported in nuclear physics \cite{bohr} and, since then, it has proved crucial to understand a whole series of major properties in nuclei such as the odd-even staggering in binding energies and the presence of an energy gap in the spectrum of even-even nuclei \cite{brink,ring}. In recent years, in conjunction with an increased experimental exploration of nuclei far from stability, a revival of interest in pairing has been observed due to its fundamental support to the theoretical description of 
loosely bound nuclear systems \cite{dean,brink}.

In a mean-field treatment of nuclei, pairing  mimics the short range part of the 
effective nucleon-nucleon interaction. Its basic action on nucleons is that of 
scattering them between pairs of time-conjugate single-particle levels. 
In its simplest form, namely assuming this scattering to be independent of the levels involved,
the pairing Hamiltonian reads as
\begin{equation}
H=\sum_k\epsilon_kN_k-g\sum_{kk'}P^{\dag}_kP_{k'},
\label{1}
\end{equation}
where $P^{\dag}_k$ creates a pair of particles in time-conjugate states on the level $k$ of the mean field, $N_k$ counts the number of particles on this level, $\epsilon_k$ is the associate energy and $g$ the constant pairing strength. In spite of being highly schematic, this Hamiltonian has been applied to various fields of physics ranging from nuclear structure to condensed matter,  where its applications include
not only macroscopic superconducting systems but also finite size microscopic systems 
such as ultrasmall metallic grains \cite{delft}. 

The Hamiltonian (\ref{1}) is commonly treated in the BCS approximation \cite{bcs} or in the more refined Hartree-Fock-Bogoliubov (HFB) approach \cite{ring}. As it is well known, these methods, while suited to describe macroscopic systems, being  exact in the thermodynamical limit \cite{bursill}, show some limitations when applied to systems with a limited number of particles such as nuclei. This is mainly due to fact that they make use of a  wave function that is not an eigenstate of the number operator. 

In a series of papers in the 60's, Richardson \cite{richa1,richa2,richa3}, also in collaboration with Sherman \cite{richa}, studied the pairing Hamiltonian (\ref{1}) and showed that the corresponding eigenvalue problem can be solved exactly in a semi-analytical way. More precisely, it was shown that, for an even number of particles $(2N)$, the exact eigenstates of the Hamiltonian (1) can be written as a product of distinct, collective pairs of particles built on time-conjugate states. These pairs can be either real or complex, with reference to their  mixing amplitudes, depending on the value of the strength $g$. Complex pairs, in particular, always occur in a complex-conjugate form.
In  \cite{samba1}, it was shown that a product of two such complex-conjugate pairs actually forms a four-body correlated structure with real mixing amplitudes (quartet). The exact ground state of the Hamiltonian (\ref{1}) can therefore be formulated in an equivalent form as a product of these quartets. Quartets are expected to be all distinct from one another and, for odd values of $N$, they are joined in by an extra collective pair also characterized by real amplitudes. Such an exact and (semi)analytic representation of the ground state in terms of quartets holds, however, only for the constant pairing Hamiltonian (1). Extending this formalism of quartets to a general pairing Hamiltonian and investigating how much this new representation improves upon more traditional descriptions of pairing are the basic motivations of the work presented here.

Quartets will be constructed via an iterative variational approach based on diagonalizations of the Hamiltonian in restricted model spaces. A similar technique, but based on pairs, has been employed in a previous work on pairing \cite{samba2} and a comparison with its results will be discussed below. 
Each diagonalization is meant to update a quartet while  driving the ground state towards its minimum. Whenever needed, an extra real collective pair, also determined variationally,  will be added to these quartets. 
As a testing ground for the present approach, we will examine the Sn isotopes with the valence neutrons
in the 50-82 neutron shell. Comparisons will be carried out with the exact solutions as well as with more traditional approaches to pairing. The analysis will not be limited to the ground state only but it will concern low-lying excited states as well. As a final step, a simplified representation of the ground state as a product of identical quartets will be formulated and compared with its direct pair analogue, i.e. the  particle-number projected-BCS ground state \cite{ring}.

The paper is organized as follows. In section 2, we  review the Richardson approach to the constant
pairing Hamiltonian and point out the emergence of quartets in its ground state. In section 3, we  describe the procedure which has been developed to construct the quartets (and the extra pair) in the case of a general pairing Hamiltonian. In section 4, we present the results relative to the ground state while, in section 5, we  extend the
quartet formalism to excited states. In section 6, we discuss the representation of the ground
state as a product of identical quartets and, finally, in section 7, we  summarize the results 
and draw the conclusions. 

\section{Quartets in the Richardson formalism}
This section will be devoted to an analysis of the constant pairing Hamiltonian
\begin{equation}
H=\sum_i\epsilon_i{\cal N}_i-g\sum_{ii'}L^{\dag}_iL_{i'},
\label{12}
\end{equation}
where
\begin{equation}
{\cal N}_i=\sum_{m_i}a^\dag_{im_i}a_{im_i},~~~~
L^{\dag}_i=\sum_{m_i> 0}P^\dag_{im_i},~~~~
P^\dag_{im_i}=a^\dag_{im_i}{\tilde{a}}^\dag_{im_i}.
\label{13}
\end{equation}
The operator $a^\dag_{im_i}$ creates a particle on the single particle level $i$, with energy $\epsilon_i$ and degeneracy $\Omega_i$, ${\tilde{a}}^\dag_{im_i}$ being the time reversed operator. In the case of pairing in a spherically symmetric mean field, $i\equiv n_i,l_i,j_i$ (adopting the standard notation), $\Omega_i =2j_i+1$ and ${\tilde{a}}^\dag_{im_i}=(-1)^{j_i-m_i}a^\dag_{i-m_i}$, $m_i$ being the projection of $j_i$. In this case, the operator $L^{\dag}_i$ creates a pair of particles on the level $i$ with total angular momentum $J=0$. In the absence of spherical symmetry, $L^{\dag}_i$ simply creates a pair of particles with opposite projections of the intrinsic spin on the doubly degenerate level $i$ ($m_i=\pm$ and $\Omega_i =2$ in this case).

By further defining 
\begin{equation}
N_{im_i}=a^\dag_{im_i}a_{im_i}+a^\dag_{i-m_i}a_{i-m_i},
\label{14}
\end{equation}
the Hamiltonian (\ref{12}) reduces to
\begin{equation}
H=\sum_{i,m_i>0}\epsilon_iN_{im_i}-g\sum_{i,m_i>0}\sum_{i',m_{i'}>0}
P^\dag_{im_i}P_{i'm_{i'}}
\end{equation}
which, assuming the simplified notation $k\equiv i,m_i>0$, is nothing but the Hamiltonian (\ref{1}). The index $k$ will be herewith assumed to vary in the interval $(1,\Omega )$.

The eigenvalue problem relative to the Hamiltonian (\ref{12}) can be solved by applying the Richardson formalism \cite{richa1,richa2,richa3,richa}. One finds that, for a system of $N$ pairs and by neglecting the presence of unpaired particles in the model space (these do not partecipate in the pair-scattering generated by the Hamiltonian (\ref{1}) \cite{richa}), any eigenstate can be formulated as the pair product state
\begin{equation}
|\Psi\rangle =\prod^{N}_{\nu=1}B^{\dag}_{\nu}|0\rangle,~~~~~
B^{\dag}_{\nu}=\sum^{\Omega}_{k=1}\frac{1}{2\epsilon_k-e_{\nu}}P^{\dag}_k
\label{2}
\end{equation}
if the $N$ parameters $e_\nu$ (the so-called pair energies) are roots of
the set of $N$ coupled non-linear equations
\begin{equation}
1-\sum^{\Omega}_{k=1}\frac{g}{2\epsilon_k-e_\nu}+
\sum^{N}_{\rho=1(\rho\neq \nu)}\frac{2g}{e_\rho-e_\nu}=0.
\label{3}
\end{equation}
The eigenvalue $E^{(\Psi )}$ associated with $|\Psi\rangle$ is just the sum
of the corresponding pair energies, i.e.
\begin{equation}
E^{(\Psi )}=\sum^{N}_{\nu=1}e_\nu. 
\label{222}
\end{equation}

Solving numerically the equations (\ref{3}) meets  some technical difficulties. Indeed,
by varying the strength $g$, it can happen that
two (real) pair energies $e_\nu$ become equal thus giving
rise to a singularity. When this occurs, these energies
turn from real into complex-conjugate pairs. 
It is in order to handle only real quantities 
that Richardson introduced some new variables $\xi_\lambda$ and
$\eta_{\lambda}$ \cite{richa}. 
In the case of the ground state and for an even number $N$ of pairs, these are defined such that 
\begin{equation}
e_{2\lambda -1}=\xi_\lambda -i\eta_\lambda ,~~~~
e_{2\lambda}=\xi_\lambda +i\eta_{\lambda}~~~~
(\lambda =1,2,\cdot\cdot\cdot ,N/2).
\label{bec1}
\end{equation}
$\xi_\lambda$ is assumed to be always real while
$\eta_\lambda$ can be either pure imaginary (corresponding to real and distinct
$e_{2\lambda -1}$ and $e_{2\lambda}$) or real (corresponding to complex-conjugate pair energies). Once the transformation (\ref{bec1}) is
made, the equations (\ref{3}) depend only on the real quantities 
$\xi_\lambda$ and $\eta^2_\lambda$ which can therefore be calculated numerically
(although at the cost of a further change of variables \cite{richa}). With increasing $g$, the quantities $\eta^2_\lambda$ turn one by one from negative to positive values so generating the complex-conjugate pairs.
The case of odd $N$
can be described in the same formalism of equation (\ref{bec1}) by assuming that one pair energy is real and the remaining $N-1$ pair energies occur in complex-conjugate 
pairs \cite{richa}. 

In the new variables (9) the ground state for a system of an even number of pairs can be written
as a product of complex-conjugate collective pairs
\begin{equation}
|\Psi_{gs}\rangle =\prod^{N/2}_{\lambda =1}
B^{\dag}_{2\lambda -1}B^{\dag}_{2\lambda}|0\rangle
\label{bec2}.
\end{equation}
The complex-conjugate form (\ref{bec1}) guarantees that the product 
$B^{\dag}_{2\lambda -1}B^{\dag}_{2\lambda}$ can be set into a real form no matter what $g$. 
Indeed, one finds that \cite{samba1}
\begin{equation}
B^{\dag}_{2\lambda -1}B^{\dag}_{2\lambda}=(\Gamma^\dag_\lambda )^2+\eta^2_\lambda
(\Theta^\dag_\lambda)^2
\label{bec3},
\end{equation}
where the pairs
\begin{eqnarray}
&&\Gamma^\dag_\lambda =\sum^{\Omega}_{k=1}\frac{2\epsilon_k-\xi_\lambda}
{(2\epsilon_k-\xi_\lambda )^2 +\eta^2_\lambda}P^\dag_k
\equiv\sum^{\Omega}_{k=1}\gamma^{(\lambda )}_kP^\dag_k,\label{aqw}\\
&&\Theta^\dag_\lambda =\sum^{\Omega}_{k=1}\frac{1}
{(2\epsilon_k-\xi_\lambda )^2 +\eta^2_\lambda}P^\dag_k
\equiv\sum^{\Omega}_{k=1}\theta^{(\lambda )}_kP^\dag_k,\label{aqwe}
\end{eqnarray}
have amplitudes $\gamma^{(\lambda )}_k$ and $\theta^{(\lambda )}_k$  depending only
on the real coefficients $\xi_\lambda$ and $\eta^2_\lambda$. More explicitly, one has      \begin{equation}
B^{\dag}_{2\lambda -1}B^{\dag}_{2\lambda}=\sum^{\Omega}_{k,k'=1}
(\gamma^{(\lambda )}_k\gamma^{(\lambda )}_{k' }+\eta^2_\lambda
\theta^{(\lambda )}_k\theta^{(\lambda )}_{k' })P^\dag_kP^\dag_{k'}
\label{bec4}.
\end{equation}
This expression holds true both for real and complex-conjugate pairs. The product $B^{\dag}_{2\lambda -1}B^{\dag}_{2\lambda}$ can therefore be always represented as  a real superposition of products of two pairs of particles in time-reversed states. In the following, we will refer to a four-body correlated structure having these general properties as a quartet. The ground state of the pairing Hamiltonian (\ref{12}) for a system with an even number $N$ of pairs consequently acquires the form of a product of quartets: these are, in general, partly amenable to the product of two real collective pairs and partly associated with the product of two complex-conjugate pairs. For a system with odd $N$, instead, the ground state can always be reduced to a product of quartets and an extra real collective pair.

\section{Quartets for a general pairing Hamiltonian}
The analysis of section 2 cannot be extended to the more general pairing Hamiltonian
\begin{equation}
H=\sum_i\epsilon_i{\cal N}_i-\sum_{ii'}g_{ii'}L^{\dag}_iL_{i'}
\label{112}
\end{equation}
since no analytical solution of the eigenvalue problem is known in this case. It is nevertheless possible to carry out an approximate treatment of the eigenstates of Hamiltonian (15) in terms of quartets 
by resorting to numerical methods. In this section, we will show how this has been realized.

We begin by considering a system with an even number $N$ of pairs and we
assume a ground state of the form
\begin{equation}
|\Psi_{gs}\rangle =\prod^{N/2}_{\nu=1}Q^{\dag}_{\nu}|0\rangle,~~~~~
Q^{\dag}_{\nu}=\sum^{(1,\Omega )}_{k<k'}q^{(\nu )}_{kk'}P^{\dag}_kP^{\dag}_{k'}
\label{200}.
\end{equation}
The operator $Q^{\dag}_{\nu}$ is a generalization of the operator (\ref{bec4}) and creates a quartet with real amplitudes $q^{(\nu )}_{kk'}$ to be determined. $|\Psi_{gs}\rangle$ is therefore a product of quartets which are, in principle, all distinct from one another.
In order to search for the most appropriate $q^{(\nu )}_{kk'}$'s we make use of an iterative variational procedure that resorts to diagonalizations of the Hamiltonian in spaces of a rather limited size. This procedure draws inspiration from an analogous technique previously developed for a treatment of pairing correlations in terms of a set of independent pairs \cite{samba2} and works as follows. Let us suppose that, at a given stage of the iterative process, one knows the state (\ref{200}) and let us construct the space 
\begin{equation}
F^{(\rho)}=\Biggl\{ P^\dag_kP^\dag_{k'} \prod^{N/2}_{\nu=1(\nu\neq \rho)}Q^\dag_\nu |0\rangle 
\Biggr\}_{1\leq k<k'\leq\Omega}.
\label{5}
\end{equation}
The states of $F^{(\rho)}$ are generated by acting with the operators $P^\dag_kP^\dag_{k'}$ $(k<k')$ on the product of all the quartets
$Q^\dag_\nu$ but the $\rho$-th one. The dimension of each space (\ref{5})   is therefore $\Omega (\Omega -1)/2$ and one can form at most $N/2$ such spaces. By diagonalizing the Hamiltonian in $F^{(\rho)}$ and searching for the lowest eigenstate, one constructs the state
\begin{equation}
|\Psi_{gs}^{(new)}\rangle =Q^{\dag (new)}_\rho\prod^{N/2}_{\nu=1(\nu\neq \rho)}Q^\dag_\nu |0\rangle 
\label{6}.
\end{equation}
This differs from $|\Psi_{gs}\rangle$ only for the new quartet $Q^{\dag (new)}_\rho$ and its
energy is by construction lower than (or, at worst, equal to) that of 
$|\Psi_{gs}\rangle$. As a result of this operation, then, the quartet $Q^{\dag (new)}_\rho$ has updated $Q^{\dag}_\rho$ while all the other quartets have remained unchanged. At the same time the energy of  $|\Psi_{gs}\rangle$ has been driven towards its minimum. Performing a series of diagonalizations of $H$ in $F^{(\rho)}$ for all possible $\rho$ values $(1\leq\rho\leq N/2)$ exhausts what we define an iterative cycle. At the end of a cycle all the quartets $Q^\dag_\nu$ have been updated and a new cycle can then start. The sequence of iterative cycles goes on until the difference between the ground state energy at the end of two successive cycles becomes vanishingly small.

The case of an odd number $N$ of pairs proceeds along the same path but one now assumes that the ground state is formed by $(N-1)/2$ quartets and one extra pair. One has, then,
\begin{equation}
|\Psi_{gs}\rangle =B^{\dag}\prod^{(N-1)/2}_{\nu=1}Q^{\dag}_{\nu}|0\rangle,~~~~~
B^{\dag}=\sum^{\Omega}_{k=1}\beta_{k}P^{\dag}_k
\label{201},
\end{equation}
the quartets $Q^{\dag}_{\nu}$ being defined as in equation (\ref{200}) and the amplitudes $\beta_k$ being supposed to be real. In this case, the iterative cycle consists of two different steps. In the first one, one performs a series of diagonalizations in the spaces
\begin{equation}
R^{(\rho)}=\Biggl\{ P^\dag_kP^\dag_{k'} B^{\dag}\prod^{(N-1)/2}_{\nu=1(\nu\neq \rho)}
Q^\dag_\nu |0\rangle 
\Biggr\}_{1\leq k<k'\leq\Omega}
\label{202}
\end{equation}
for all possible $\rho$ values. These diagonalizations update the $(N-1)/2$ quartets while leaving unchanged the pair. Once these have been completed, as a second step, one performs a diagonalization in the space
\begin{equation}
S=\Biggl\{ P^\dag_k \prod^{(N-1)/2}_{\nu=1}Q^\dag_\nu |0\rangle 
\Biggr\}_{1\leq k\leq\Omega}
\label{203}
\end{equation}
which updates the pair $B^\dag$ with the quartets acting as spectators. Also in this case, the sequence of iterative cycles goes on up to the convergence of the ground state energy.

\section{Ground state results}
As an application of the procedure just described, we have considered a case of pairing in a spherically symmetric mean field. This application has concerned, in particular, the Sn isotopes with the valence neutrons
in the 50-82 neutron shell. As customary for these nuclei, the model space has been restricted to the five neutron orbitals $g_{7/2}$, $d_{5/2}$, $d_{3/2}$, $s_{1/2}$, and $h_{11/2}$. Single-particle energies $\epsilon_j$ and pairing strengths $g_{ii'}$ have been taken from the work of Zelevinsky and Volya \cite{zele} and are the same used in a previous analysis of pairing \cite{samba2}. Within this model space, exact calculations have been performed by resorting to the method of  \cite{volya} which is based on a classification of the many-body configurations within the seniority scheme \cite{rac1,rac2,kerm}  and on the use of the SU(2) quasispin algebra.

The quartets $Q^{\dag}_{\nu}$ of equation (\ref{200}) are defined in an uncoupled formalism and so they are not characterized by a good angular momentum $J$. The same is true for the pair $B^{\dag}$ of equation (\ref{201}). In spite of that, $J=0$ pair and quartets can be generated quite naturally by proceeding step by step for increasing values of $N$. For $N=1$, the diagonalization of the Hamiltonian in the space $S$ (\ref{203}) generates a $J=0$ pair as ground state. For $N=2$, a similar diagonalization in $F^{(1)}$ gives rise to a $J=0$ quartet. Moving to $N=3$ (a system which, in our formalism, is described as a product of one pair and one quartet), it is sufficient to use the previously generated $J=0$ pair and quartet as an initial ansatz to generate, at each step of the iterative cycle, either a $J=0$ pair or a $J=0$ quartet. The same mechanism holds true going to larger $N$'s. 
We notice that, as a result of the degeneracies of the Hamiltonian, the coefficients $q^{(\nu )}_{kk'}$ that are generated are independent of the angular momentum  projections and so the quartets (\ref{200}) acquire the simple form
\begin{equation}
Q^{\dag}_{\nu}=\sum_{i\leq i'}q^{(\nu )}_{i,i'}/(1+\delta_{i,i'})L^{\dag}_iL^{\dag}_{i'}
\label{23},
\end{equation}
with the operators $L^{\dag}_i$ (\ref{13}) creating a $J=0$ pair  on the orbital $i$. These $J=0$ quartets are therefore a linear combination of products of two uncorrelated $J=0$ pairs.

In figure 1, we have plotted the relative error in the ground state correlation energy (i.e., the energy of the ground state relative to the lowest uncorrelated many-body configuration) as a function of the number $N$ of pairs. Results obtained with the procedure of section 3 are shown at various levels of approximation, each level corresponding to a different number of iterative cycles (indicated by the number next to each line). Already at the lowest level of approximation (1 cycle), the approach is seen to provide a remarkably good agreement with the exact results. This agreement progressively improves with increasing the number of iterative cycles up to becoming basically exact. 

It is of interest to compare these results with those which can be obtained in a formalism of 
independent $J=0$ pairs. The dot-dashed line in figure 1 refers to a calculation in which the pairing ground state has been assumed to be a product of real, collective, distinct, $J=0$ pairs \cite{samba2}. The approach based on quartets appears to be considerably more effective than the corresponding one with $J=0$ pairs.

In figure 2, we have plotted the root mean square values of the relative errors in the occupation numbers, i.e. the quantity $\sigma (oc)=\sqrt{(\sum^5_{i=1}\Delta^2_i(n))/5}$  with
$\Delta_i(n)=(n^{(exact)}_i-n^{(appr)}_i)/n^{(exact)}_i$,  the index $i$ running over the 5 orbitals of the shell 50-82 and $n_i$ being the corresponding occupation numbers. In figure 3, one finds the analogous rms values for the pair transfer matrix elements, i.e.
$\sigma (tr)=\sqrt{(\sum^5_{i=1}\Delta^2_i(t))/5}$  where
$\Delta_i(t)=(t^{(exact)}_i-t^{(appr)}_i)/t^{(exact)}_i$ and 
$t_i=|\langle\Psi (N)|L^\dag_i|\Psi(N-1)\rangle|$. The notation is the same as in figure 1, i.e., 
the dot-dashed line in both figures refers to the calculation with $J=0$ pairs while the remaining lines show the results obtained in the formalism with quartets at various levels of approximation. The results of these figures confirm the analysis of figure 1. 

\section{Excited states}
The formalism of quartets illustrated in section 3 for the ground state also lends itself to a straightforward extension to excited states. To such a purpose, once the procedure relative to the ground state has been completed, we proceed by diagonalizing the Hamiltonian either in the space
\begin{equation}
G^{(E)}=\Biggl\{ P^\dag_kP^\dag_{k'} \prod^{N/2}_{\nu=1(\nu\neq \rho)}Q^\dag_\nu |0\rangle 
\Biggr\}_{1\leq k<k'\leq\Omega,\rho=1,2,\cdot\cdot\cdot ,N/2}
\label{50}
\end{equation}
or in the space
\begin{equation}
\fl G^{(O)}=\Biggl\{ P^\dag_kP^\dag_{k'} B^{\dag}\prod^{(N-1)/2}_{\nu=1(\nu\neq \rho)}
Q^\dag_\nu |0\rangle , P^\dag_k \prod^{(N-1)/2}_{\nu=1}Q^\dag_\nu |0\rangle
\Biggr\}_{1\leq k<k'\leq\Omega,\rho=1,2,\cdot\cdot\cdot ,(N-1)/2}
\label{51},
\end{equation}
depending on the number of pairs $N$: $G^{(E)}$ is employed for systems with even $N$ (i.e., in the presence of quartets only), $G^{(O)}$ when $N$ is odd. $G^{(E)}$ includes at once all the states of the spaces $F^{(\rho)}$ (\ref{5}). These are, more specifically, all the states that can be formed by replacing, one at a time, each quartet of the ground state with all its basic components $P^\dag_kP^\dag_{k'}$. $G^{(O)}$ is formed, instead, by all the states of the spaces $R^{(\rho )}$ (\ref{202}) plus those of $S$ (\ref{203}). In addition to those states which are obtained by replacing each quartet with its components $P^\dag_kP^\dag_{k'}$, $G^{(O)}$ therefore includes the states which result from the replacement of the pair $B^\dag$ with its components $P^\dag_k$. One should notice that, as a result of the diagonalization of the Hamiltonian in these spaces, the excited states are no longer a simple product of quartets (and a pair, in case) as for the ground state but they are rather a linear combination of these products. These diagonalizations also lead to a rederivation  of the ground state: this fact guarantees the orthogonality between this state and the excited ones.

In tables 1 and 2, we show some examples of excitation spectra that are generated with this procedure.
We adopt the same formalism used in the derivation of the exact eigenstates (section 4) and classify the approximate excited states according to the seniority quantum number $v$ \cite{rac1,rac2,kerm}, that is to say the number of nucleons not pairwise coupled to angular momentum zero. The energies shown in table 1 are those of the lowest 10 $v=0$ excited states in two middle shell Sn isotopes, namely $^{116}$Sn and $^{118}$Sn. All these states are therefore characterized by a $J=0$ angular momentum. Table 2 shows, instead, the energies of the lowest 10 $v=2$  states still for $^{116}$Sn and $^{118}$Sn. In this case, each energy is usually associated with different $J\neq 0$ degenerate eigenstates. The agreement between exact and approximate results is pretty good.

Some comments on the internal structure of the excited states just discussed are in order. Quartets forming $v=0$ eigenstates are always of the type (\ref{23}) ($J=0$ quartets). In $v=2$ eigenstates, instead, each component is characterized by one $J\neq 0$ quartet of the type
\begin{equation}
[a^\dag_{j}a^\dag_{j}]^J_0(\sum_i c_iL^{\dag}_i)
\label{35},
\end{equation}
the remaining ones being still $J=0$ quartets.
This quartet is a product of a collective $J=0$ pair and an uncorrelated $J\neq 0$ pair. As it can be seen from the above expression, following the general form (\ref{200}) of the quartets, the latter pair is characterized by two nucleons with the same angular momentum $j$.
We finally remark that these quartets are such that one cannot 
form $v=4$ (or higher) states with good total angular momentum. The present formalism is therefore not well suited to describe these states.

\section{The ground state as a product of identical quartets}
The description of the ground state that has been pursued so far is based on quartets which are 
all distinct from one another. Treating systems of like particles (neutrons in the case just analyzed of Sn isotopes), 
quartets have been defined in terms of these particles only.
In an attempt to describe proton-neutron pairing, use has often been made in literature 
of $\alpha$-like quartets, namely four-body structures composed of two neutrons and two protons. 
In one of the earliest works on this subject \cite{flowers}, $\alpha$-like quartets have been introduced in a BCS-type wavefunction. 
Later on, more sophisticated wavefunctions have been proposed, still of BCS-type, which include the contribution of quartet correlations
to the proton-neutron pairing (e.g., \cite{chasman,senkov}). As well known, however, the use of BCS-type wavefunctions has 
the drawback of not conserving neither the particle number nor the isospin of the nucleus. Quite recently, an accurate description 
of the proton-neutron isovector pairing in  nuclei has been obtained with a quartet model  which does not violate any symmetry 
of the Hamiltonian \cite{nicu1,nicu2}. In this approach, the nuclear ground state is simply described as a product of 
identical $\alpha$-like quartets with angular momentum $J=0$ and isospin $T=0$.
The present analysis offers the interesting opportunity to investigate, in a case of pairing among like particles, 
how well a description of the ground state as a product of identical quartets can approximate a description in which 
quartets are left free to be distinct from one another. To such a purpose, we have therefore compared the results 
of section 4 with those which are obtained by representing the ground state as 
\begin{equation}
(Q^ \dag_\nu)^{N/2}|0\rangle
\label{36},
\end{equation}
in cases of even $N$, or as
\begin{equation}
B^ \dag(Q^ \dag_\nu)^{(N-1)/2}|0\rangle
\label{37},
\end{equation}
in cases of odd $N$. In these expressions, we have kept the same definitions of  
$Q^ \dag_\nu$ and $B^\dag$ given in equations (\ref{200}) and (\ref{201}), respectively. The amplitudes $q^{(\nu )}_{kk'}$ and $\beta_k$ have been determined by minimizing the ground state energy.
In figure 4, we compare the relative errors in the ground state correlation energy obtained in the approaches with identical quartets (triangles) and with independent quartets (squares). For completeness, in the same figure (circles), we also show the same quantity calculated by approximating the ground state as a product of identical pairs $B^\dag$ only, that is to say in the well known particle-number projected-BCS approximation (PBCS) \cite{ring}. Clear differences can be observed between these three approaches, the corresponding errors often differing from one another by something like two orders of magnitude. The worst approximation appears to be the PBCS one
(it should be noticed, however, that even in this case the relative errors remain confined within a quite acceptable range). The approach with identical quartets performs definitely better than the corresponding one with pairs and shows itself as a very good approximation of the exact ground state.

\section{Summary and conclusions}
In this work, we have provided a description of the low-lying spectrum of a general pairing Hamiltonian for a finite system of like particles in a formalism of quartets. These are, in general terms, real superpositions of products of two pairs in time-reversed states.
Quartets have been first formulated in an analytical form in the case of a constant pairing 
Hamiltonian. This concept has  been then extended to a general pairing Hamiltonian. To such a purpose, a numerical procedure has been proposed which is based on an iterative sequence of diagonalizations of the Hamiltonian. This procedure has been applied to the Sn isotopes with the valence neutrons in the 50-82 
neutron shell.

Two major features of this approach are worthy being emphasized: it does not violate any symmetry of the Hamiltonian and it reduces the original shell model diagonalization, usually involving very large spaces (up to 6x$10^8$ basis states, in the uncoupled formalism, in the case of Sn isotopes \cite{volya}), to a sequence of diagonalizations in reduced spaces of only a few hundreds of states. As a major result, we have found that a product of distinct $J=0$ quartets (and, in case, an extra $J=0$ pair) can provide a basically exact representation of the pairing ground state. The quartet formalism has also been extended to excited states 
showing  that it describes well the seniority 0 and 2 eigenstates. 
Comparisons with a traditional approach to pairing like the particle-number projected-BCS approach, in which
the ground state is supposed to be a product of identical $J=0$ pairs, or 
with a more sophisticated approximation based on distinct $J=0$ pairs,
have evidenced a  much higher quality of the approximation based on quartets. 
Finally, we have shown that a simplified description of the ground state as a product of identical quartets provides an excellent approximation of the exact ground state.

The formalism of quartets developed in this work has shown to be quite successful in describing 
like-particle pairing in a realistic context. In order to be extended to the treatment of more 
general systems, such as protons and neutrons interacting through isovector and isoscalar pairing forces, 
this formalism needs to be further developed by a redefinition of the basic operators
forming the quartets. This development is the scope of future studies.

\ack
 N.S acknowledges the support from UEFISCDI of Romanian Ministry
 of Education and Research through Grant IDEI  No 57

\newpage
\begin{table}
\caption{Comparison between exact and approximate energies of the lowest 10 $v=0$ excited states in $^{116}$Sn and $^{118}$Sn. Approximate results have been obtained with the procedure of section 5. All values are in MeV.}
\begin{indented}
\lineup
\item[]\begin{tabular}{ccccc}
\br
\multicolumn{2}{c}{$^{116}$Sn}& \multicolumn{1}{c}{} &\multicolumn{2}{c}{$^{118}$Sn}\\
\cline{1-2} \cline{4-5}
 ~~~exact~~~ & approximate & ~~~~~~ & ~~~exact~~~ & approximate\\
\br
 2.3571 & 2.3571 & ~~~~~~ & 2.3311 & 2.3311 \\ 
 3.0115 & 3.0116 & ~~~~~~ & 2.8941 & 2.8942 \\ 
 3.2495 & 3.2496 & ~~~~~~ & 3.6361 & 3.6361 \\ 
 4.0229 & 4.0229 & ~~~~~~ & 4.5070 & 4.5070 \\ 
 4.7795 & 4.7807 & ~~~~~~ & 4.5569 & 4.5577 \\ 
 5.1974  & 5.2010 & ~~~~~~ & 5.4049  & 5.4069 \\
 5.9277  & 5.9320 & ~~~~~~ & 5.5611  & 5.5620 \\
 6.0754  & 6.0790 & ~~~~~~ & 5.7648  & 5.7673 \\
 6.4924  & 6.4968 & ~~~~~~ & 6.6544  & 6.6556 \\
 6.5964  & 6.5984 & ~~~~~~ & 6.9403  & 6.9429 \\
\br
\end{tabular}
\end{indented}
\end{table}

\begin{table}
\caption{As in table 1, for $v=2$ states.}
\begin{indented}
\lineup
\item[]\begin{tabular}{ccccc}
\br
\multicolumn{2}{c}{$^{116}$Sn}& \multicolumn{1}{c}{} &\multicolumn{2}{c}{$^{118}$Sn}\\
\cline{1-2} \cline{4-5}
 ~~~exact~~~ & approximate & ~~~~~~ & ~~~exact~~~ & approximate\\
\br
 2.9832 & 2.9839 & ~~~~~~ & 2.6957 & 2.6958 \\ 
 3.2691 & 3.2694 & ~~~~~~ & 3.0218 & 3.0218 \\ 
 3.8398 & 3.8403 & ~~~~~~ & 4.2904 & 4.2905 \\ 
 4.5816 & 4.5817 & ~~~~~~ & 4.5047 & 4.5088 \\ 
 5.1152 & 5.1263 & ~~~~~~ & 4.6708 & 4.6720 \\ 
 5.2917 & 5.2973 & ~~~~~~ & 5.0742 & 5.0742 \\
 5.7828 & 5.8034 & ~~~~~~ & 5.1998 & 5.2109 \\
 5.8861 & 5.8907 & ~~~~~~ & 5.6664 & 5.6704 \\
 6.0401 & 6.0456 & ~~~~~~ & 5.8045 & 5.8047 \\
 6.2093 & 6.2439 & ~~~~~~ & 6.3618 & 6.3651 \\
\br
\end{tabular}
\end{indented}
\end{table}

\begin{figure}
\includegraphics{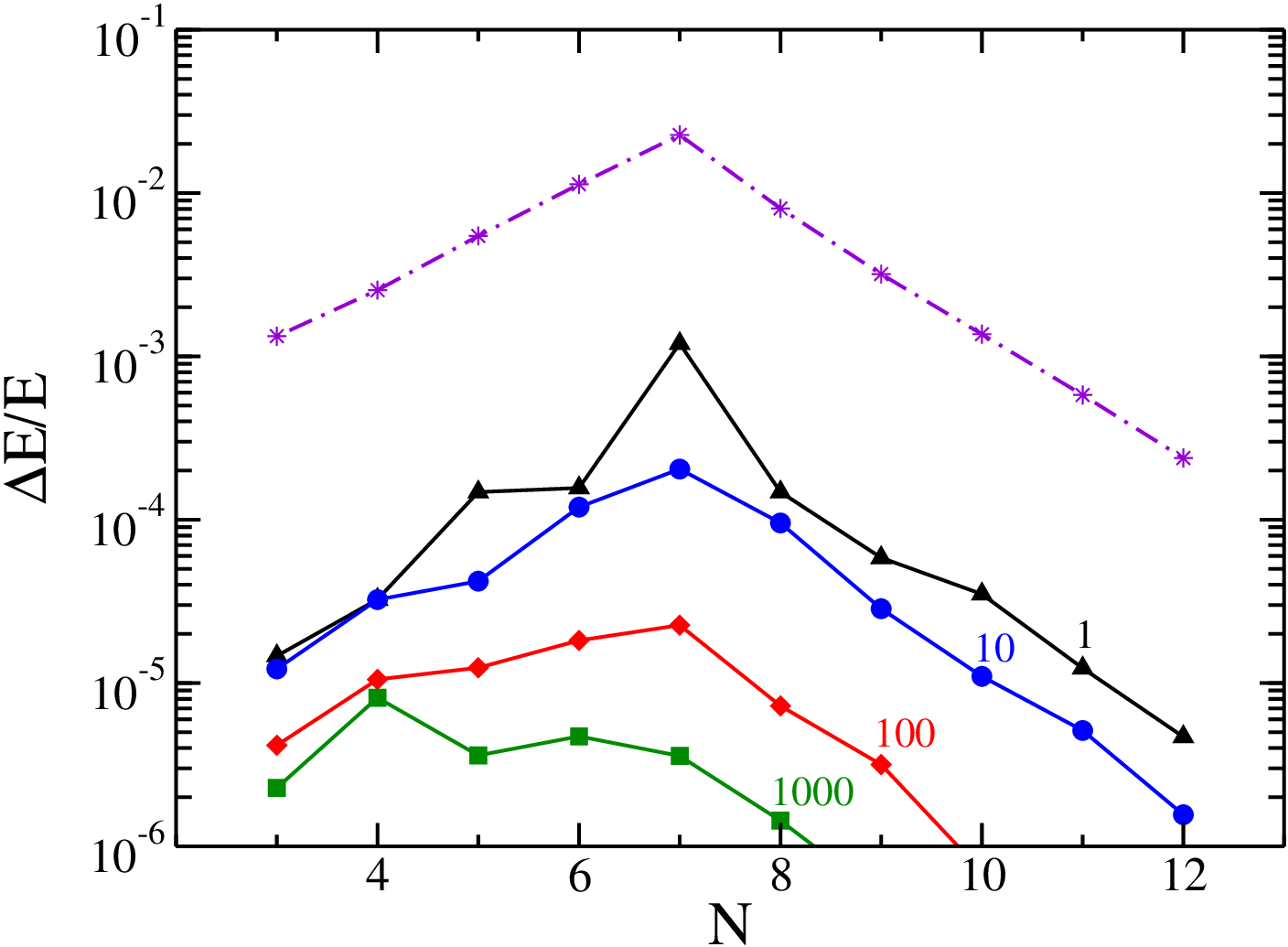}
\mbox{}\\[19cm]
\caption{(Colour online) Relative errors in the ground state correlation energy of Sn isotopes with $2N$ active neutrons. Solid lines refer to the procedure of section 3 at various levels of approximations: the number next to each line indicates how many iterative cycles have been carried out in each calculation (see text). The dot-dashed line shows the same quantity calculated with the approach of  \cite{samba2} based on independent $J=0$ pairs .} 
\end{figure}

\begin{figure}
\includegraphics{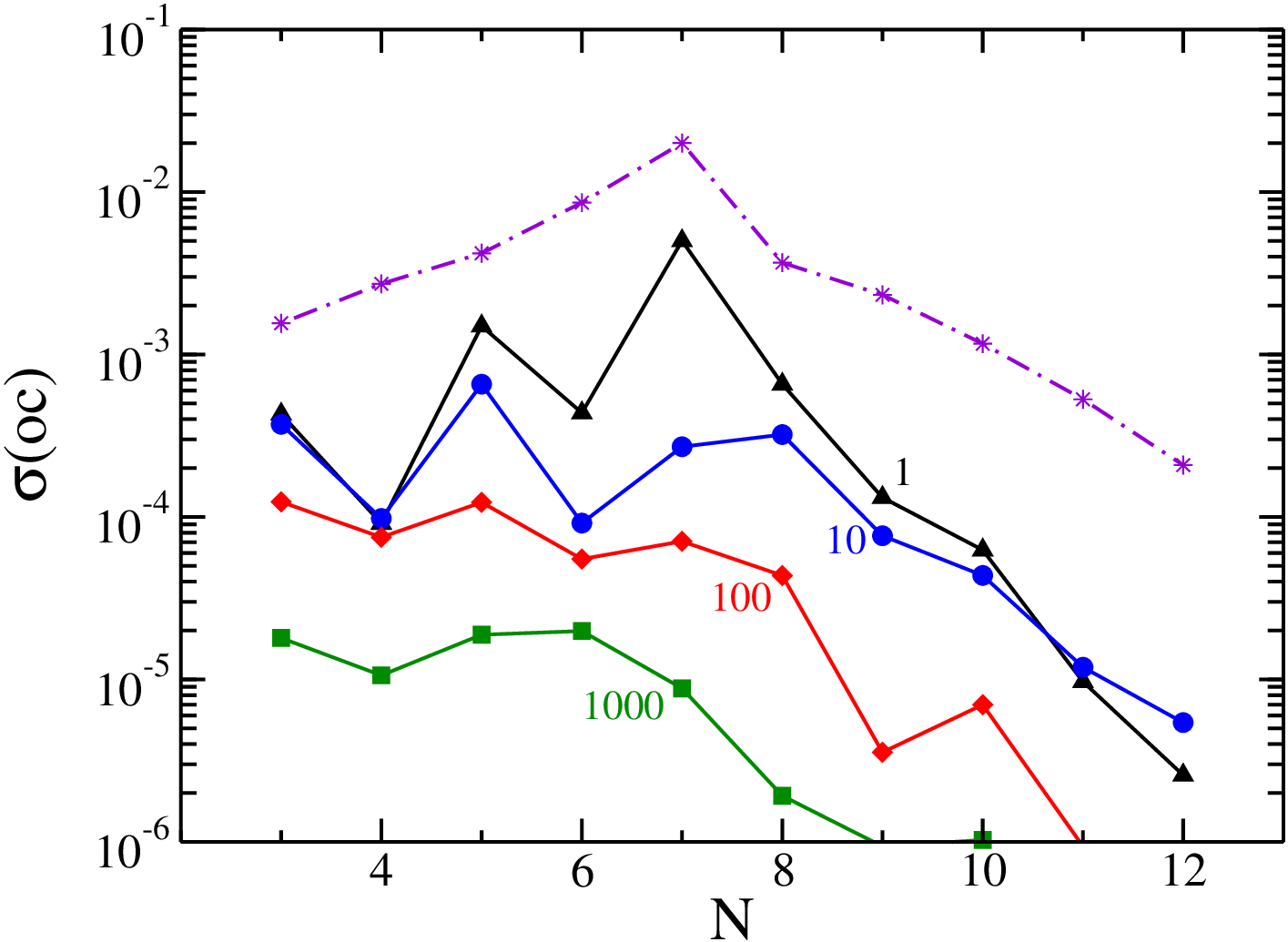}
\mbox{}\\[19cm]
\caption{(Colour online) Root mean square values of the relative errors in the occupation numbers calculated  for Sn isotopes with $2N$ active neutrons (see text). Symbols are the same as in figure 1.}
\end{figure}

\begin{figure}
\includegraphics{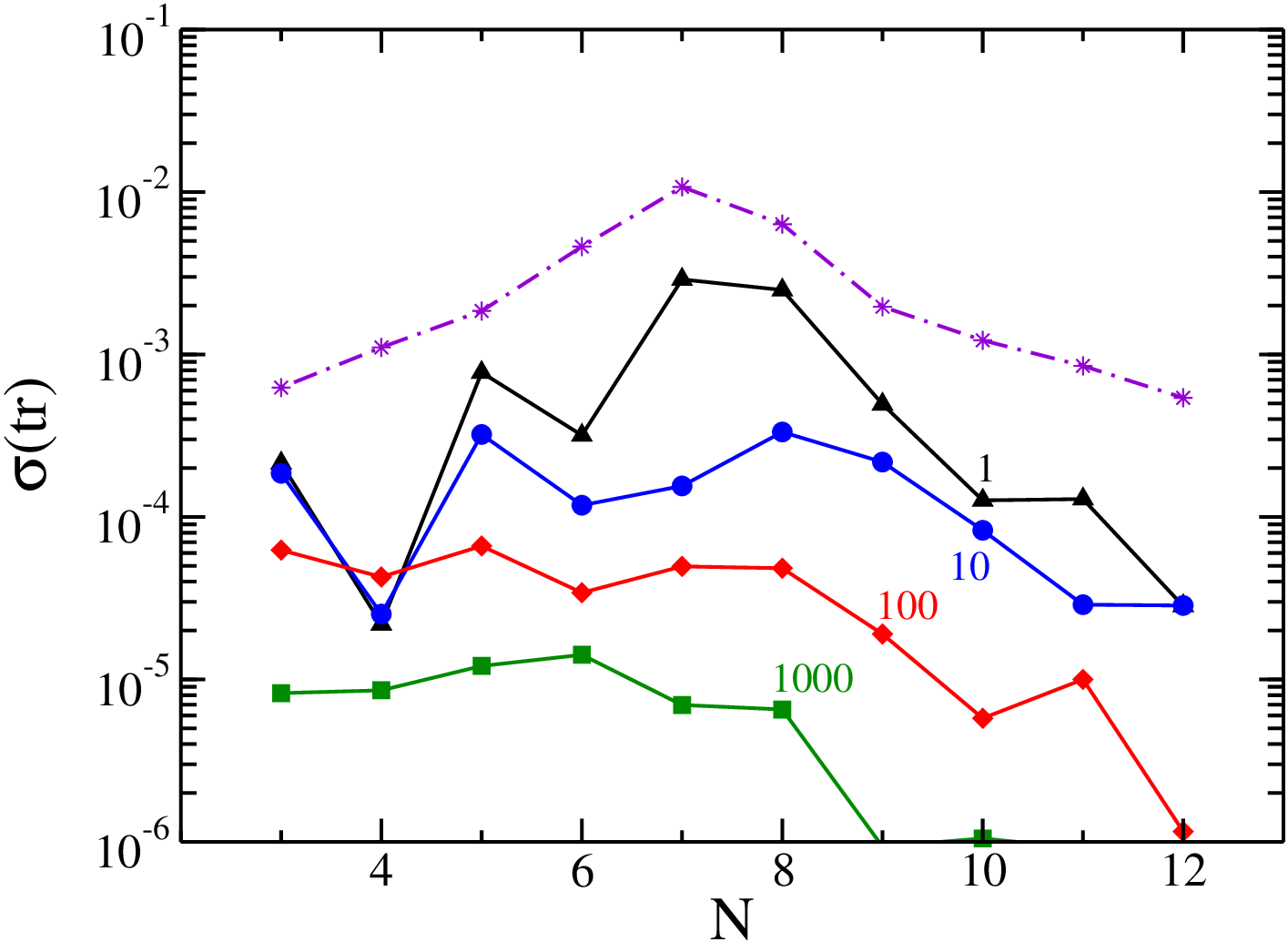}
\mbox{}\\[19cm]
\caption{(Colour online) Root mean square values of the relative errors in the pair transfer matrix elements calculated  for Sn isotopes with $2N$ active neutrons (see text). Symbols are the same as in figure 1.}
\end{figure}

\begin{figure}
\includegraphics{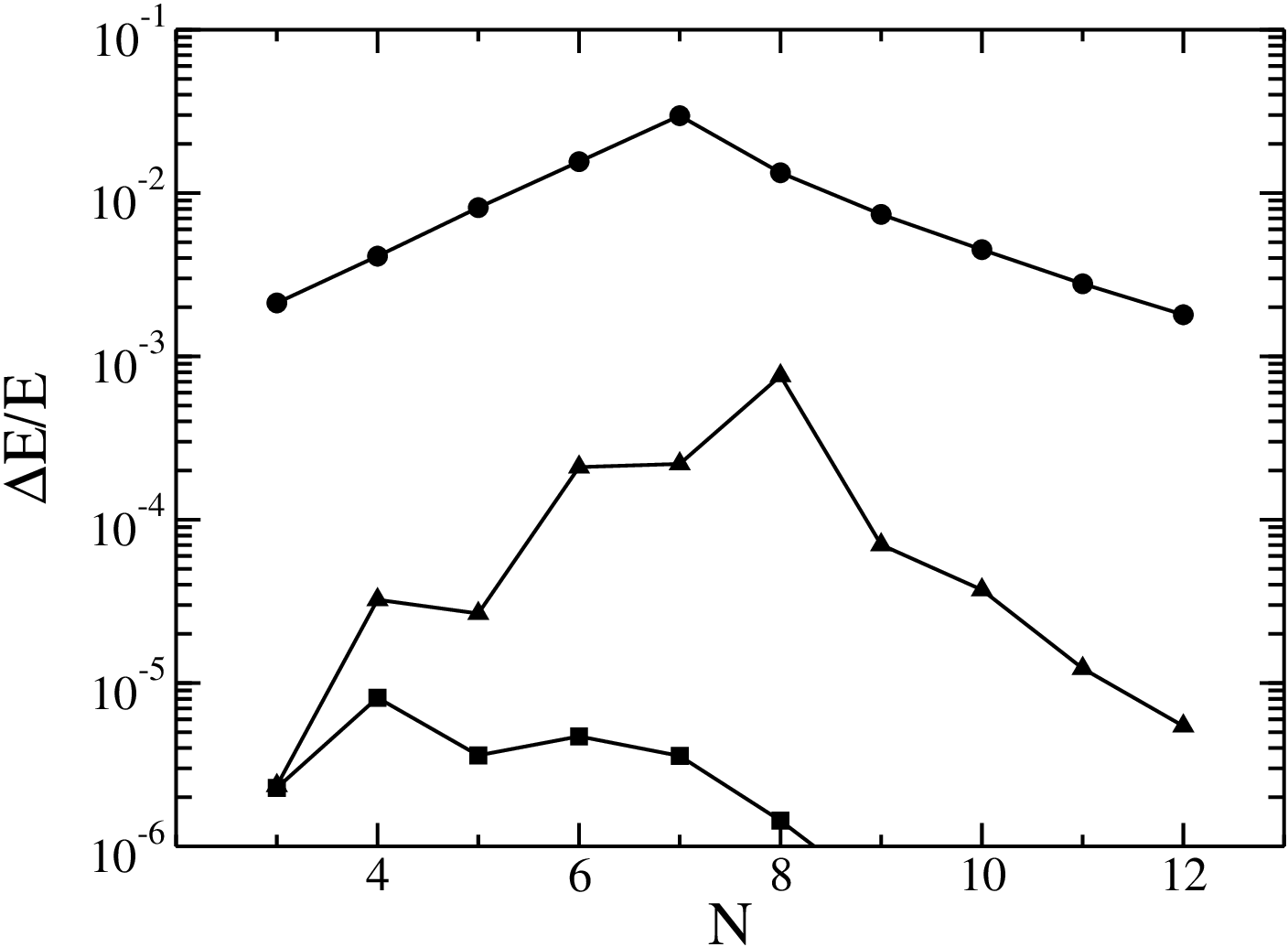}
\mbox{}\\[19cm]
\caption{Relative errors in the ground state correlation energy of Sn isotopes with $2N$ active neutrons. The line labeled with triangles shows the results obtained within the approach of section 6 with identical quartets. The line labeled with circles refers to the calculations in the PBCS scheme (identical pairs). The line labeled with squares shows the best results obtained with the procedure of section 3.}
\end{figure}

\end{document}